\documentclass[twocolumn,showpacs,preprintnumbers,draftclsnofoot]{revtex4}
\usepackage{amsmath}
\usepackage{amssymb}
\usepackage{graphicx}
\usepackage{dcolumn}
\usepackage{bm}
\usepackage{amssymb}
\usepackage{graphicx}
\usepackage{dcolumn}
\usepackage{bm}

\begin{document}

\title{ Magically strained bilayer graphene with flat bands   }
\author{Ma Luo\footnote{Corresponding author:luom28@mail.sysu.edu.cn} }
\affiliation{The State Key Laboratory of Optoelectronic Materials and Technologies \\
School of Physics\\
Sun Yat-Sen University, Guangzhou, 510275, P.R. China}

\begin{abstract}

Twist bilayer graphenes with magical angle have nearly flat band, which become strongly correlated electron systems. Herein, we propose another system based on strained bilayer graphene that have flat band at the intrinsic Fermi level. The top and bottom layers are uniaxially stretched along different directions. When the strength and directions of the strain satisfy certain condition, the periodical lattices of the two layers are commensurate to each other. The regions with AA, AB and BA stacking arrange in a triangular lattice. With magical strain, the bands around the intrinsic Fermi level are nearly flat and have large gap from the other bands. This system could provide more feasible platform for graphene-based integrated electronic system with superconductivity.

\end{abstract}

 \maketitle

\section{Introduction}

Narrow band strongly correlated electronic systems exhibit many exotic physical features, including superconductor, Mott insulator and spin liquid. The systems could be briefly described by Hubbard model, which is a tight binding model with hopping strength $t$ and on-site Coulomb interaction $U$. The systems with $U/t>5$ could be consider to be strongly correlated. For realistic graphene, this ratio is $U/t=1.6$ \cite{Schuler13}, so that graphene is not strongly correlated system. However, a twisted bilayer graphene \cite{LopesdosSantos07,EJMele10,Shallcross11,EJMele11,LopesdosSantos12,GuyTrambly16,Mingjian18,Ramires18,Adriana18,YoungWooChoi18,JianKang18,Yndurain19} with magical twisting angle have isolated flat band \cite{Morell10,Pilkyung13,Hirofumi17,Nguyen17,Angeli18,Naik18} near to the intrinsic Fermi level. The effective tight binding model for the flat band \cite{Venderbos18,Koshino18,Yuan18} exhibit the feature of strongly correlated electronic system because $U/t^{\prime}>5$. Recent experiments confirm that the phase diagram of the systems include superconductor and Mott insulator phase \cite{YuanCao181,YuanCao182}, or topological nontrivial phase \cite{Spanton18,SSSunku18,JiaBinQiao18,Shengqiang18}. A lot of recent theoretical result have been devoted to explain the superconductivity and other novel physics in this system \cite{ChengCheng18,Sherkunov18,Rademaker18,YuPingLin18,Peltonen18,HoiChun18,YingSu18,Kennes18,Fengcheng18,Cenke18,Hiroki18,Gonzalez19,Hejazi19}.

We proposed that a bilayer graphene with magical strain could also have isolated flat band. The top and bottom layers are uniaxially stretched along different directions. The strain could be induced by substrates or external pulling forces. We present the conditions to have commensurate lattice for the two layers. Under the suitable strain, the AA stacking region is arranged into the triangular lattice, which is similar to that for the twisted bilayer graphene. The band structure under magical strain are calculated. Previoius studies of strained bilayer graphenes only have partially flat bands \cite{SeonMyeongChoi10,VanderDonck16,MVogl17}. Recent theoretical studies show that the partially flat bands in the graphene with periodical strain could host superconductivity \cite{FengXu18}. Because our system have nearly completely flat band, the superconductivity could be more easily induced.

The article is organized as following: In section II, the lattice structure of the magically strained bilayer graphene is presence. In section III, the band structure given by the tight binding model is presented. In section IV, the conclusion is given.

\section{The Lattice Structure}

The lattice structure of a single layer graphene is defined by the primitive translation vectors of the primitive unit cell, which are $\mathbf{a}_{1}=\sqrt{3}a_{c}\hat{x}$ and $\mathbf{a}_{2}=\frac{\sqrt{3}}{2}a_{c}\hat{x}+\frac{3}{2}a_{c}\hat{y}$, with $a_{c}=0.142$ nm being the bond length. The zigzag edge of the graphene is along x axis. The locations of the A sublattice are $\mathbf{r}_{A}(n,m)=n\mathbf{a}_{1}+m\mathbf{a}_{2}$, and the locations of the B sublattice are $\mathbf{r}_{B}(n,m)=n\mathbf{a}_{1}+m\mathbf{a}_{2}+\frac{\sqrt{3}}{2}a_{c}\hat{x}+\frac{1}{2}a_{c}\hat{y}$, with $n$ and $m$ being integers. Under uniaxial strain along direction $\hat{T}=\cos(\theta)\hat{x}+\sin(\theta)\hat{y}$ with strength $\varepsilon$, the location of each lattice site is transferred into $\mathbf{r}_{A(B)}^{\theta,\epsilon}=(\mathbf{I}+\mathbf{\epsilon})\cdot\mathbf{r}_{A(B)}$, where $\mathbf{I}$ is a two-by-two unit matrix and $\mathbf{\epsilon}$ is the strain tensor. The strain tensor is given as \cite{Pereira09}
\begin{equation}
\mathbf{\epsilon}=\varepsilon\begin{bmatrix} \cos^{2}\theta-\sigma\sin^{2}\theta & (1+\sigma)\cos\theta\sin\theta \\ (1+\sigma)\cos\theta\sin\theta & \sin^{2}\theta-\sigma\cos^{2}\theta \end{bmatrix}
\end{equation}
where $\sigma=0.165$ is the Poisson＊s ratio of graphene. After the strain, the primitive translation vectors are transferred into $\mathbf{a}_{1(2)}^{\theta,\epsilon}=(\mathbf{I}+\mathbf{\epsilon})\cdot\mathbf{a}_{1(2)}$. In consequence, the first Brillouin zone becomes distorted hexagon.
If the strength of the strain satisfies $\varepsilon<0.23$, the band structure remains gapless. The K and K$^{\prime}$ Dirac points with linear band touching move away from the K and K$^{\prime}$ points of the first Brillouin zone. With $\varepsilon=0.23$, the K and K$^{\prime}$ Dirac points merge. With $\varepsilon>0.23$, bulk band gap opens \cite{Pereira09}.

\begin{figure}[tbp]
\scalebox{0.46}{\includegraphics{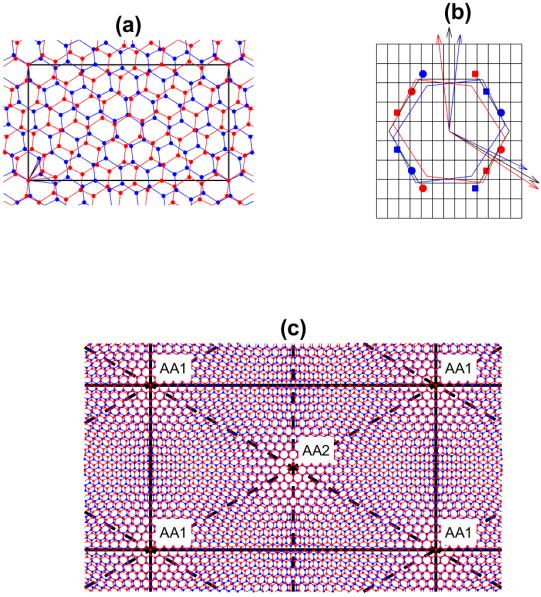}}
\caption{ (a) The lattice structure of the bilayer graphene with strain parameter $N_{\theta}=0$, $q=2$ and $p=1$. The lattice sites and bonds of the top and bottom graphene layers are plotted as blue and red, respectively. (b) The reciprocal lattice vectors and Brillouin zone of unstrained graphene(black), top strained graphene layer(blue) and bottom strained graphene layer(red). The rectangular grid is the Brillouin zones of the superlattice. The circle and square points are the K and K$^{\prime}$ Dirac points with linear band touching for the corresponding single layer graphene, respectively. (c) The lattice structure of the bilayer graphene with strain parameter $N_{\theta}=0$, $q=10$ and $p=1$. \label{fig1}}
\end{figure}

The magical strained bilayer graphene is consisted of an AA stacking bilayer graphene, whose top and bottom layers are strained with the same strength but along $\theta$ and $-\theta$ direction, respectively. We designate the primitive translation vectors of the top layer as $\mathbf{a}_{1(2)}^{\theta,\epsilon}:=\mathbf{a}_{1(2)}^{T}$ and those of the bottom layer as $\mathbf{a}_{1(2)}^{-\theta,\epsilon}:=\mathbf{a}_{1(2)}^{B}$. In addition, we define another translation vectors as $\mathbf{a}_{1^{\prime}}^{T(B)}=\mathbf{a}_{2}^{T(B)}-\mathbf{a}_{1}^{T(B)}$. The translation vectors of the superlattice is given by the linear combination of the primitive translation vectors of the top layer as
\begin{eqnarray}
&&\mathbf{a}_{1}^{R}=q\mathbf{a}_{1^{\prime}}^{T}+(q+p)\mathbf{a}_{2}^{T} \nonumber \\
&&\mathbf{a}_{2}^{R}=-q^{\prime}\mathbf{a}_{1^{\prime}}^{T}+(q^{\prime}+p)\mathbf{a}_{2}^{T} \label{superlattice1}
\end{eqnarray}
, where $q$ and $p$ are positive integer. The conditions that the supperlattice is rectangular lattice are $\mathbf{a}_{1}^{R}\cdot\hat{x}=0$ and $\mathbf{a}_{2}^{R}\cdot\hat{y}=0$, which gives the condition $q^{\prime}=3q+p-\sqrt{3}p\frac{\cos2\theta}{\sin2\theta}$. The same translation vectors of the superlattice should be given by the similar linear combination of the primitive translation vectors of the bottom layer as
\begin{eqnarray}
&&\mathbf{a}_{1}^{R}=(q+p)\mathbf{a}_{1^{\prime}}^{B}+q\mathbf{a}_{2}^{B} \nonumber \\
&&\mathbf{a}_{2}^{R}=(-q^{\prime}-p)\mathbf{a}_{1^{\prime}}^{B}+q^{\prime}\mathbf{a}_{2}^{B} \label{superlattice2}
\end{eqnarray}
The condition that the superlattice is commensurate with the lattice of top and bottom layer is $q^{\prime}$ being integer. As a result, the angle of the strain direction should satisfies the condition
\begin{equation}
\theta=\frac{1}{2}\tan^{-1}\frac{\sqrt{3}}{N_{\theta}}
\end{equation}
, where $N_{\theta}$ is a non-negative integer, and the strength of the strain is given as
\begin{equation}
\varepsilon=\frac{-p}{p\cos^{2}\theta-\sqrt{3}(p+2q)(1+\sigma)\cos\theta\sin\theta-p\sigma\sin^{2}\theta}
\end{equation}
The ratio between the length of the two translation vectors is $|\mathbf{a}_{2}^{R}|/|\mathbf{a}_{1}^{R}|=\sqrt{3}$. An example of such bilayer graphene with $N_{\theta}=0$, $q=2$ and $p=1$ is plotted in Fig. \ref{fig1}(a). The reciprocal lattice vectors and the first Brillouin zone of the top and bottom graphene layers are plotted in Fig. \ref{fig1}(b). The first Brillouin zone of the two graphene layers are distorted in opposite way. The Dirac points with linear band touching of the two layers (in the absence of the inter-layer coupling) are related by the symmetric operation $y\rightarrow-y$, or by the combined symmetric operation $\{x\rightarrow-x,K\leftrightarrow K^{\prime}\}$. The effect of the inter-layer coupling between two states with the same $k_{x}$ or $k_{y}$ from the two layers might be weak. The band structure of the bilayer graphene is the mix of two Dirac cone with linear dispersion with weak mixing strength, so that flat dispersion along $\hat{y}$ direction could appears. Another example of such bilayer graphene with $N_{\theta}=0$, $q=10$ and $p=1$ is plotted in Fig. \ref{fig1}(c). The regions with AA, AB and BA stacking are clearly identified. The AA stacking regions locate at the four corners and the middle of the unit cell of the superlattice. All AA stacking regions are arranged in a triangular lattice. However, the lattice structures of the AA1 stacking region and the AA2 stacking region are a little different, so that the whole system is not strict triangular lattice of AA stacking regions as that in twist bilayer graphene. The AB and BA stacking regions also arrange in the triangular lattice. Both type of stacking regions also have two subtypes, designated as AB1(BA1) and AB2(BA2).

In this article, we focus on the superlattice defined in this section with $N_{\theta}\in\{1\sim20\}$, $q\in\{3\sim23\}$ and $p=1$. The band structures of a few cases with optimal figure of merit are plotted and discussed in the next section. Many other type of superlattice could be constructed by varying $p$ or the form of Eq. (\ref{superlattice1}) and (\ref{superlattice2}). The supercells could contain more than two AA stacking regions and too many lattice sites in the supercell, which makes the numerical simulation infeasible. As an example, another superlattice with triangular lattice symmetric is presented in the Appendix.

\section{The Band Structure}

The band structure of the magically strained bilayer graphene in the superlattice can be calculated by tight binding model. The Hamiltonian is given as
\begin{equation}
H=-\sum_{\langle i,j\rangle}{t_{i,j}c_{i}^{\dag}c_{j}}+c.c.
\end{equation}
, where $c_{i}^{(\dag)}$ is the annihilation (creation) operator of the electron at the $i$-th lattice cite, $t_{i,j}$ is the hopping parameter between the $i$-th and $j$-th lattice cites. The summation index cover the pairs of lattice cites $\langle i,j\rangle$, whose distance between each other is smaller than $5a_{c}$. The detail expression of $t_{i,j}$ could be found in multiple references \cite{Pilkyung13,Nguyen17,Adriana18,JianKang18}, which includes the intra-layer and inter-layer hopping parameter. Applying the Bloch boundary condition of the superlattice, the band structure of the rectangular lattice could be calculated. We found that two types of isolated narrow band exist under certain straining parameters. For the first type of band structure, eight bands around the intrinsic Fermi level are sticked together, which have finite band gap from the higher and lower bands. For the second type of band structure, two bands below the intrinsic Fermi level are nearly flat, which have finite band gap from the higher and lower bands. We define the figure of merit for the narrow bands as
\begin{equation}
M=\frac{min(E_{gap}^{c},E_{gap}^{v})}{E_{w}}
\end{equation}
where $E_{gap}^{c(v)}$ is the gap from the conduction(valence) band, and $E_{w}$ is the band width of the narrow bands.

\begin{figure}[tbp]
\scalebox{0.5}{\includegraphics{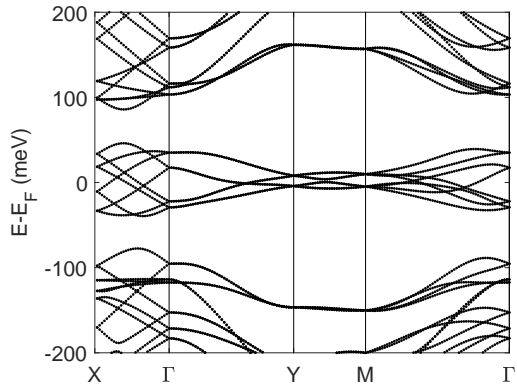}}
\caption{ Band structure of the magically strained bilayer graphene with $N_{\theta}=6$, $q=18$ and $p=1$. \label{fig2}}
\end{figure}

The magically strained bilayer graphene with the first type of isolated narrow bands exist for the straining parameter $N_{\theta}\in\{4\sim8\}$. The optimal figure of merit is $0.4447$, for the systems with $N_{\theta}=6$ and $q=18$. The band structure is plotted in Fig. \ref{fig2}. The band width of the eight bands around the intrinsic Fermi level is $85.6$ meV, and the gaps from the higher and lower bands are $39.7$ meV and $38.0$ meV, respectively. The strength of the strain is $\varepsilon=0.1067$.

\begin{figure}[tbp]
\scalebox{0.5}{\includegraphics{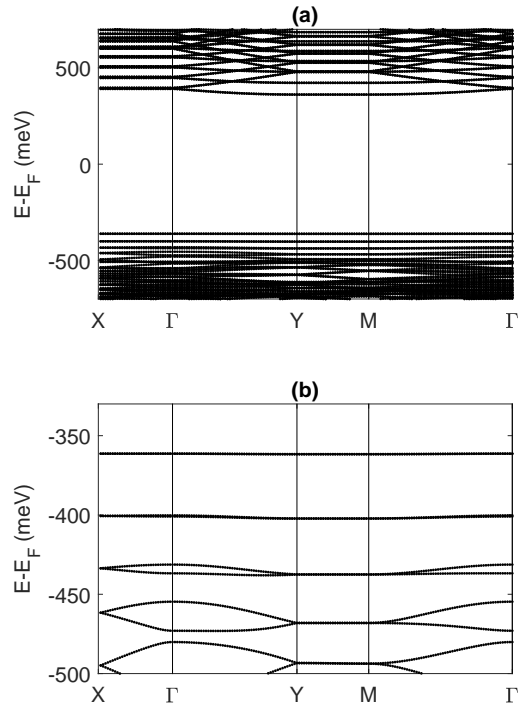}}
\caption{ Band structure of the magically strained bilayer graphene with $N_{\theta}=18$, $q=22$ and $p=1$. (a) and (b) are the same band structure, except that the y-scale in (b) is zoomed in around the flat band below the intrinsic Fermi level.  \label{fig3}}
\end{figure}

The magically strained bilayer graphene with the second type of isolated narrow bands exist for the straining parameter $N_{\theta}>9$. For larger $N_{\theta}$, the figure of merit could be higher, but the strength of the strain is also larger. For realistic materials, we are only interested in the systems with strain smaller than $0.3$. With this constrain, the optimal system have straining parameters as $N_{\theta}=18$ and $q=22$, whose figure of merit is $108$. The band structure is plotted in Fig. \ref{fig3}. The band width of the two bands below the intrinsic Fermi level is $0.36$ meV, and the gaps from the higher and lower bands are $722.7$ meV and $38.6$ meV, respectively. The strength of the strain is $\varepsilon=0.2984$. The result shows that the third and fourth bands below the intrinsic Fermi level are also nearly flat. The bands above the intrinsic Fermi level are partially flat. Specifically, along the $k_{x}$ direction with $k_{y}$ being fixed, the bands around the intrinsic Fermi level have ultra flat dispersion; along the $k_{y}$ direction, the bands are slightly dispersive. Inspection of the wave functions of the flat bands show that the electrons in these states are localized at the AA stacking regions, i.e. the localized Wannier orbitals at the AA stacking region. This feature is similar to that of twist bilayer graphene with magical angle \cite{Trambly10}. The features of the band structure implies that the localized Wannier orbitals at the AA1 stacking regions do not couple with that at the AA2 stacking regions. The localized Wannier orbitals only weakly couple to the other localized Wannier orbitals at the nearest neighbor AA stacking regions of the same type( AA1 to AA1, or AA2 to AA2). With finite doping, the Fermi surface are nearly straight lines along $k_{x}$ direction, exhibiting the feature of Fermi surface nesting. In the presence of Hubbard interaction and hole doping, the charge density wave and pair density wave could coexist with chiral
d-wave superconductivity \cite{FengXu18}.

\section{conclusion}

In conclusion, by stretching the top and bottom layer of a bilayer graphene with the same strength of strain along different directions, $Moir\acute{e}$ pattern of AA stacking regions could be generated. The AA stacking regions are arranged in a triangular lattice, but the lattice structure of the regions at two sublattices are slightly different. With magical straining parameter, the bands below the intrinsic Fermi level become ultra-flat. The quantum states in these ultra-flat bands are highly localized at the AA stacking regions.

\section{Appendix}

\begin{figure}[tbp]
\scalebox{0.33}{\includegraphics{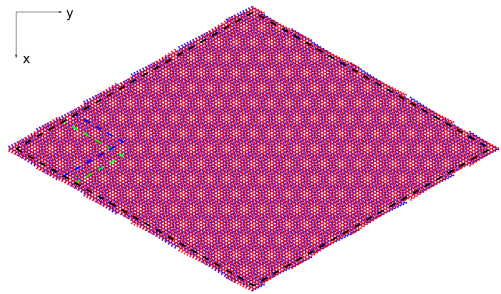}}
\caption{ The lattice structure of the magically strained bilayer graphene with triangular lattice.  \label{figA}}
\end{figure}

The variant of Eq. (\ref{superlattice1}) and (\ref{superlattice2}) could construct different types of superlattice. Herein, we present an example that the superlattice is triangular lattice. Assuming that the translation vectors of the supercell have the same direction as the primitive translation vectors of the unstrained graphene, the relation between the translation vectors of the supercell and the primitive translation vectors of the unit cell of the top strained graphene single layer is given as
\begin{eqnarray}
&&q\mathbf{a}_{1^{\prime}}^{T}+\mathbf{a}_{2}^{T}=r_{1}\mathbf{a}_{1} \nonumber \\
&&-\mathbf{a}_{1^{\prime}}^{T}+q\mathbf{a}_{2}^{T}=r_{2}\mathbf{a}_{2} \label{superlatticeT1}
\end{eqnarray}
, where $q$ is integer, $r_{1}$ and $r_{2}$ are real number. In order to satisfy this condition, the angle and strength of the strain is given as
\begin{equation}
\theta=\pm\frac{1}{2}[\tan^{-1}(-\frac{\sqrt{3}q}{2})+\pi]
\end{equation}
and
\begin{equation}
\varepsilon=\frac{2\sqrt{4+3q^{2}}}{\sqrt{4+3q^{2}}(\sigma-1)+q^{2}(\sigma+1)}
\end{equation}
Inserting the straining parameter into Eq. (\ref{superlatticeT1}), the length of the translation vectors of the supercell is given as
\begin{equation}
r_{1}=\frac{(q+2)(1+q^{2})(\sigma+1)}{\sqrt{4+3q^{2}}(\sigma-1)+q^{2}(\sigma+1)}
\end{equation}
and
\begin{equation}
r_{2}=\frac{(q-2)(1+q^{2})(\sigma+1)}{\sqrt{4+3q^{2}}(\sigma-1)+q^{2}(\sigma+1)}
\end{equation}
Because $r_{1}\ne r_{2}$, the superlattice of the top strained graphene single layer does not form triangular lattice. The bottom graphene single layer is strained along $-\theta$ direction with the same strength, so that the translation vectors of the supercell is obtained by exchanging $r_{1}$ and $r_{2}$. Defining $r_{s}=lcm(q+2,q-2)$ as the least common multiple of $q+2$ and $q-2$, the translation vectors of the supercell of the combined bilayer of the top and bottom strained graphene are given as $\mathbf{a}_{1}^{Tri}=\frac{r_{s}r_{1}}{q+2}\mathbf{a}_{1}$ and $\mathbf{a}_{2}^{Tri}=\frac{r_{s}r_{2}}{q-2}\mathbf{a}_{2}$. Thus, the supercell of the combined bilayer form a triangular superlattice. An example with $q=18$ is shown in Fig. \ref{figA}. The strength of the strain is $0.1778$. The supercells of the top and bottom single layer are plotted as blue and green dash lines, respectively. The supercell of the combined bilayer is plotted as black dash line. Because $r_{s}=lcm(16,20)=80$, $r_{s}/16=5$ and $r_{s}/20=4$, the supercell of the bilayer system is consisted of the repeats of the supercell of each single layer graphene along $\mathbf{a}_{1}$ and $\mathbf{a}_{2}$ for four or five times. Along $\mathbf{a}_{1}$ or $\mathbf{a}_{2}$ direction, the number of AA stacking regions within the supercell of the bilayer system is $n_{AA}=2qr_{s}/(q^{2}-4)$, so that there are $(n_{AA})^{2}$ regions with AA stacking within one supercell. All of these AA stacking regions are arranged in triangular lattice but have different lattice structure. For this example, there are $26000$ lattice site in one supercell, so that the diagonalization of the tight binding Hamiltonian is time consuming. If $q$ is larger, the difference among the lattice structure of all AA stacking regions become small. Thus, the whole system is near to a triangular lattice of AA stacking regions.

\begin{acknowledgments}
The project is supported by the National Natural Science Foundation of China (Grant:
11704419).
\end{acknowledgments}

\clearpage

\end{document}